\documentclass[preprint,12pt]{elsarticle}
\usepackage{graphicx}
\usepackage{dcolumn}
\usepackage{bm}
\usepackage{longtable}
\usepackage{subfigure}
\usepackage{amssymb}
\usepackage{amsmath}

\journal{Physics Letters B}

\begin{document}

\topmargin 0pt

\begin{frontmatter}

\title{First Study of the Negative Binomial Distribution Applied to Higher Moments of Net-charge and Net-proton Multiplicity Distributions}


\author{Terence J. Tarnowsky}
\author{Gary D. Westfall}
\address{National Superconducting Cyclotron Laboratory, Michigan State University, 640 S. Shaw Lane, East Lansing, MI 48824, USA}

\date{\today}

\begin{abstract}
A study of the first four moments (mean, variance, skewness, and kurtosis) and their products ($\kappa\sigma^{2}$ and $S\sigma$) of the net-charge and net-proton distributions in Au+Au collisions at $\sqrt{\rm s_{NN}}$ = 7.7-200 GeV from HIJING simulations has been carried out. The skewness and kurtosis and the collision volume independent products $\kappa\sigma^{2}$ and $S\sigma$ have been proposed as sensitive probes for identifying the presence of a QCD critical point. A discrete probability distribution that effectively describes the separate positively and negatively charged particle (or proton and anti-proton) multiplicity distributions is the negative binomial (or binomial) distribution (NBD/BD). The NBD/BD has been used to characterize particle production in high-energy particle and nuclear physics. Their application to the higher moments of the net-charge and net-proton distributions is examined. Differences between $\kappa\sigma^{2}$ and a statistical Poisson assumption of a factor of four (for net-charge) and 40\% (for net-protons) can be accounted for by the NBD/BD. This is the first application of the properties of the NBD/BD to describe the behavior of the higher moments of net-charge and net-proton distributions in nucleus-nucleus collisions.
\end{abstract}

\begin{keyword}


\end{keyword}

\end{frontmatter}


\section{Introduction}

Colliding heavy ions is an essential experimental tool for exploring the nature of the deconfinement phase transition. The phase transition line between hadronic and partonic matter in the phase diagram of quantum chromodynamics (QCD) is frequently drawn as a function of baryon-chemical potential ($\mu_{\rm B}$) and temperature. Heavy ion collisions at a particular center-of-mass (CMS) energy essentially probe a singular value (small range) of $\mu_{\rm B}$. Baryon-chemical potential can be changed by adjusting the CMS collision energy. The Relativistic Heavy Ion Collider (RHIC) recently embarked on a program to study the QCD phase diagram by colliding Au-nuclei at CMS collision energies ($\sqrt{\rm s_{NN}}$) of $\sqrt{\rm s_{NN}}$ = 7.7, 11.5, 19.6, 27, 39, 62.4, and 200 GeV \cite{STARBES}. This `energy scan' provides information at different values of $\mu_{\rm B}$ from $\approx$ 20-400 MeV \cite{STARBES, Cleymans}. Lattice QCD (LQCD) and experimental data suggest that if there is a phase transition at high temperatures close to $\mu_{\rm B} = $ 0, it appears to be a smooth crossover \cite{Lattice_Aoki}. Additional predictions from LQCD (dependent on the number of quark flavors and their masses) indicate the possible existence of a first order phase transition at larger values of $\mu_{\rm B}$. If this is the case, where the first order phase transition line and crossover phase transition region meet could be a critical point \cite{Lattice_CP}. 

From previous studies of the thermodynamics of phase transitions, it has been seen that a system undergoing a phase transition that passes through a critical point can exhibit enhanced fluctuations (e.g. critical opalescence) \cite{CriticalOp, CriticalOp2}. Lattice QCD studies have shown that near the crossover phase transition the fourth-order quark number susceptibilities of conserved quantities (baryon quantum number, strangeness quantum number, and charge) change rapidly \cite{Lattice_Cheng, Gavai_Gupta}. Because experiments cannot measure all produced baryons, strangeness, or charge, the higher-order moments of the following proxy distributions are measured for each conserved quantity (baryon quantum number, strangeness quantum number, and charge): net-proton, net-kaon, and net-charge, respectively. In this case, the first through fourth-order moments correspond to the mean, variance, skewness, and kurtosis, respectively. There have been estimates that the skewness ($S$) and kurtosis ($\kappa$) should be sensitive to critical fluctuations generated by a phase transition trajectory that passes through (or close to) a critical point. The higher moments are proportional to powers of the correlation length ($\xi$) with the skewness (third moment) proportional to $\xi^{4.5}$ and the kurtosis (fourth moment) proportional to $\xi^{7}$ \cite{Stephanov1}.

The moments of the (e.g. net-charge or net-proton) distributions contain trivial system volume dependencies. Constructing products and ratios of these moments can cancel the intrinsic volume dependence. Two examples of these volume-independent products are $S\sigma$ and $\kappa\sigma^{2}$. These ratios of moments can also be expressed in terms of cumulants ($C_{n}$): $S\sigma = \frac{C_{3}}{C_{2}}$ and $\kappa\sigma^{2} = \frac{C_{4}}{C_{2}}$. Quantitatively, their sensitivities to the correlation length ($\xi$) are $S\sigma\propto\xi^{2.5}$ and $\kappa\sigma^{2}\propto\xi^{5}$ \cite{Gupta}.

Several measurements of the higher-moments of net-proton, net-kaon, and net-charge distributions have been carried out \cite{STAR_Netp_PRL, STAR_Netp_QM2012, STAR_NetQK_QM2012}. These measurements are frequently compared to a baseline that assumes Poisson statistics. Deviations from this Poisson baseline have been observed. However, the deviations from the Poisson baseline are not as large as those expected from strong enhancement due to the presence of a QCD critical point. One estimate of $\kappa\sigma^{2}$ for the net-proton distribution from a QCD-based model that includes a critical point is $\kappa\sigma^{2}\approx$ 2.5, 35, or 3700 for a critical point at $\sqrt{\rm s_{NN}}$ = 200, 62.4, and 19.6 GeV, respectively \cite{STAR_Netp_PRL}. Other predictions concluded that the sign of the skewness and/or kurtosis will change at a critical point \cite{Asakawa, Stephanov2}. This behavior has been observed from lattice QCD studies that have a critical point, whose location in energy and $\mu_{\rm B}$ is dependent on parameters such as the critical temperature and the lattice spacing used \cite{Gavai_Gupta}.

Before interpreting the results from the higher-moments studies in terms of QCD critical point phenomena, the contribution from known physics must be quantified. It is well known that experimental multiplicity distributions are better described by the (negative) binomial distribution (NBD/BD) \cite{Alner_UA5_NBD, Alner_UA5_NBD2, Giovannini_NBD, Ghosh_NBD, Arneodo_EMu_NBD, Adare_PHENIX_NBD, Aamodt_ALICE_NBD}, than by a Poisson distribution. The NBD/BD and Poisson distributions are all discrete probability distributions. The Poisson distribution is a limiting case of the NBD/BD where the mean and the variance are equal. For a NBD/BD distribution, the variance is larger/smaller than the mean. There are several sources of correlation in particle production that can cause deviations from pure Poisson statistics: e.g. charge conservation, correlated production of particles from one source (e.g. strings or clusters), etc. Therefore, it is important to verify the experimental results compared to not only a Poisson distribution, but to other valid baselines that better describe the multiplicity distributions. The NBD/BD baseline is one of these cases.

\section{Analysis}

For this study, minimum bias Au+Au events at $\sqrt{\rm s_{NN}}$ = 7.7 (35 million events), 11.5 (12.2 million events), 19.6 (87.3 million events), 27 (62.4 million events), 39 (90.6 million events), 62.4 (109 million events) and 200 GeV (114 million events) were created with the default HIJING (v1.383) event generator \cite{HIJING}. Distributions of positive and negative particles and protons and anti-protons were produced. From each multiplicity distribution (positive and negative charge; proton and anti-proton) the mean value and the variance were extracted. Though the mean and variance are separately extracted from the positive and negative charged particle (or protons and anti-protons) multiplicity distributions, the higher moments of the net-charge and net-proton distributions are not automatically described by the knowledge of the mean and variance of the separate distributions. From this information, the higher order cumulants for each individual distribution could be calculated. From these, the combined cumulants of the net-charge and net-proton distributions were calculated. The additive properties of cumulants means that even order cumulants are added and odd order cumulants are subtracted (also for the NBD \cite{NBD_Cumulants_Derivation}) to calculate the required cumulants for the net-charge/proton distribution. A Poisson distribution can be fully described by the mean value $\mu$ and the cumulants $C_{n} = \mu$. For two Possion distributions with mean values $\mu_{+}$ and $\mu_{-}$, the odd cumulants of the net distribution (Skellam) are equal to $\mu_{+}-\mu_{-}$ and the even cumulants are equal to $\mu_{+}+\mu_{-}$. Ratios of the even cumulants (e.g $C_{4}/C_{2}$) are then equal to one, while $C_{3}/C_{2}$ is equal to ($\mu_{+}-\mu_{-})/(\mu_{+}+\mu_{-}$). A negative binomial distribution with mean $\mu$ and variance $\sigma^{2}$ ($\mu < \sigma^{2}$ for an NBD) is described by two parameters, $p$ and $n$ where,

\begin{align}
p=\frac{\mu}{\sigma^{2}} \\
n=\frac{\mu p}{1-p}
\end{align}

With knowledge of the mean and variance of an NBD, $p$, $n$, and the cumulants of the NBD can be calculated as,

\begin{align}
C_{1} = \frac{n(1-p)}{p} \\
C_{2} = \frac{n(1-p)}{p^{2}} \\
C_{3} = \frac{n(p-1)(p-2)}{p^{3}} \\
C_{4} = \frac{n(1-p)(6-6p+p^{2})}{p^{4}}
\end{align}

For two NBD distributions, the cumulants of each distribution are defined as either $C_{n,+}$ and $C_{n,-}$, where the odd cumulants of the net distribution are equal to $C_{n=odd} = C_{n,+} - C_{n,-}$ and the even cumulants are equal to $C_{n=even} = C_{n,+} + C_{n,-}$, from which the ratio of the cumulants can be calculated. A similar exercise can be carried out for the binomial distribution (where $\mu > \sigma^{2}$).

The centrality is defined using charged particle multiplicity in the region $0.5 < |\eta| < 1.0$. The measured charged particles and identified protons/anti-protons are restricted to the pseudorapidity range $|\eta| < 0.5$. For this measurement, this is required to prevent auto-correlations between the particles used for the measurement of the moments of the multiplicity distributions and those used for centrality determination. This method has been used in other analyses of particle correlations in pseudorapidity \cite{STAR_FB_PRL}. For net-charge studies, the transverse momentum range for particles in this analysis was $0.2 < \rm p_{T} < 2.0$ GeV/$c$ and for net-proton studies the range $\rm p_{T} > 0.4$ and $\rm p < 3.0$ GeV/$c$ was used. 

The moments are calculated in each multiplicity bin and weighted by the number of events in each bin to remove any centrality bin-width effects \cite{STAR_PtFluc_PRC, STAR_FB_PRL}.

\section{Results and Discussion}

\begin{figure}[hbtp]
\centering
\includegraphics[width=0.65\textwidth]{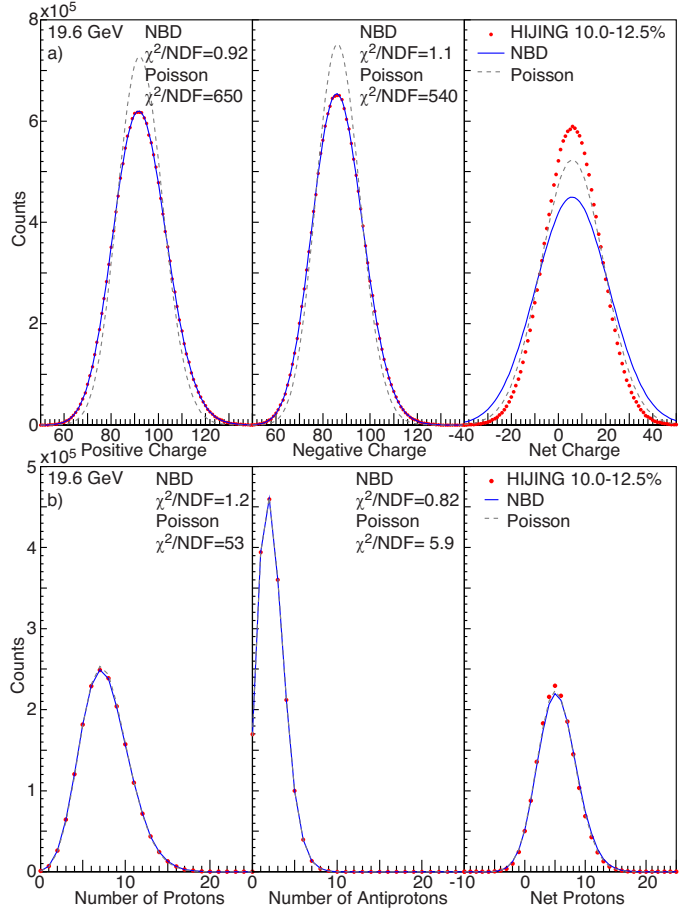}
\caption{a) Multiplicity distributions (filled circles) of positively and negatively charged particles, and the net-charge distribution (upper left, upper center, and upper right, respectively) and of b)  protons, anti-protons, and the net-proton distribution (lower left, lower center, and lower right, respectively) from $\sqrt{\rm s_{NN}}$ = 19.6 GeV Au+Au collisions in the 10-12.5\% most central collisions from HIJING. Two baseline distributions are calculated using the mean value (for Poisson) (dashed line) or the mean and variance (for negative binomial distribution (NBD)) (solid line) of the multiplicity distributions.}
\label{ChargeProtonHist}
\end{figure}

Figure \ref{ChargeProtonHist} shows the multiplicity distribution for positively and negatively charged particles and the net-charge distribution (top panels) and for protons, anti-protons, and the net-proton distribution (bottom panels) from $\sqrt{\rm s_{NN}}$ = 19.6 GeV Au+Au collisions in a representative centrality bin (10-12.5\%) from HIJING simulations. Two baseline distributions are calculated from the information in the histograms: the solid lines are a negative binomial distribution (NBD) and the dashed lines are a Poisson distribution. The only input to the Poisson curve is the mean of the multiplicity distribution, while for the NBD the mean and variance of the multiplicity distribution are required. For the positively charged particles, the $\chi^{2}/NDF$ for the Poisson is 650 and for the NBD is 0.92. For the protons, the $\chi^{2}/NDF$ for the Poisson distribution is 53, while for the NBD the $\chi^{2}/NDF$ is 1.2. Similar differences between the Poisson and NBD are observed for the negatively charged particle and anti-proton distribution. In both cases, the NBD is a better description of the data than the Poisson distribution. The same is true for negatively charged particles and anti-protons, which are treated separately.

While the positive and negative (or proton and anti-proton) multiplicity distributions are separately well described by the NBD/BD, the difference of two NBD/BD distributions to create the net-charge or net-proton distributions are not necessarily trivially described by an NBD/BD. This is seen clearly for the net-charge distribution (upper right panel) in Figure \ref{ChargeProtonHist}(a), where the NBD describes well the positively and negatively charged particle multiplicity distributions separately (Figure \ref{ChargeProtonHist} upper left and center), but does not fully describe the net-charge distribution (Figure \ref{ChargeProtonHist} upper right panel). 

\begin{figure}
\centering
\subfigure[~Net-Charge]{
\includegraphics[width=0.63\textwidth]{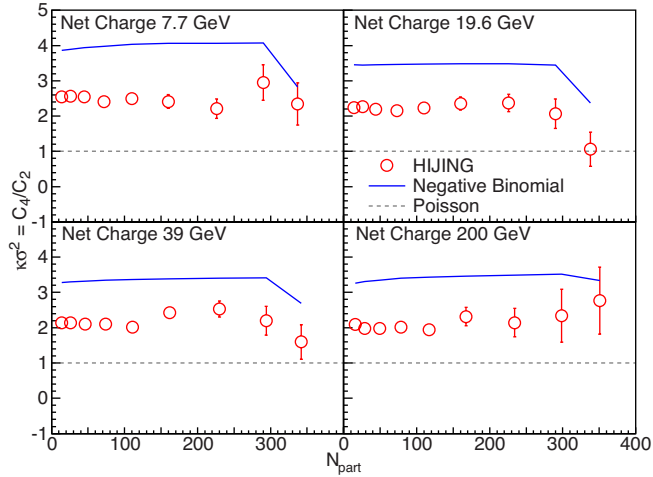}
\label{NetQKVarCent}
}
\subfigure[~Net-Protons]{
\includegraphics[width=0.63\textwidth]{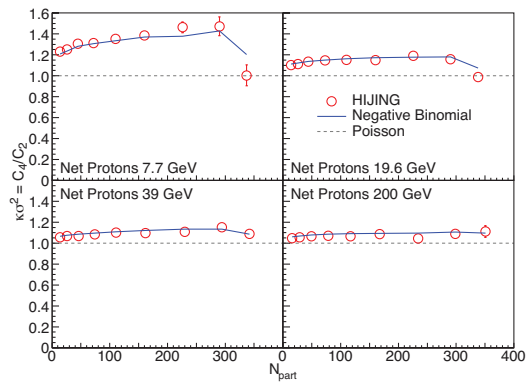}
\label{NetPKVarCent}
}
\caption{Centrality dependence of $\kappa\sigma^{2}$ ($C_{4}/C_{2}$) (open circles) for the net-charge (\ref{NetQKVarCent}) and net-proton (\ref{NetPKVarCent})distributions from HIJING simulated Au+Au collisions at $\sqrt{\rm s_{NN}}$ = 7.7, 19.6, 39, and 200 GeV. Also shown are predictions assuming the underlying multiplicity distributions are Poisson (dashed line) or a negative binomial distribution (NBD) (solid line).}
\label{CentDep}
\end{figure}

The centrality dependence of $\kappa\sigma^{2}$ ($C_{4}/C_{2}$) for the net-charge and net-proton distributions calculated from simulated Au+Au collisions at $\sqrt{\rm s_{NN}}$ = 7.7, 19.6, 39, and 200 GeV from HIJING are shown in Figures \ref{NetQKVarCent} and \ref{NetPKVarCent}. The two bins with largest $N_{\rm part}$ correspond to 0-5 and 5-10\% most central collisions, while the other seven bins are 10\% wide bins incorporating the remaining 10-80\% of the hadronic cross-section. The results from HIJING for net-charge (Figure \ref{NetQKVarCent}) demonstrate a centrality independent trend for $\kappa\sigma^{2}$, with similar average magnitudes at all energies from $\sqrt{\rm s_{NN}}$ = 7.7-200 GeV of between $\approx$2-2.5. The Poisson values for $\kappa\sigma^{2}$ are always exactly equal to one. For the NBD baseline the values are also independent of centrality, except for the most central bin, which shows an anomalous decrease. This decrease in the most central bin is also observed in the net-charge and net-proton results from HIJING (Figure \ref{NetPKVarCent}) and in the data analysis carried out by the STAR experiment \cite{STAR_Netp_PRL}. HIJING creates heavy-ion events that are a superposition of independent nucleon-nucleon collisions and does not include physics related to the deconfinement phase transition or QCD critical point. Therefore, most observables in HIJING should exhibit a smooth evolution as a function of centrality. This downward fluctuation in the most central bin seems to persist in both hadronic models and data, which indicates that it is almost certainly a consequence of the analysis method. This effect is not seen in the 5-10\% bin, which generally follows the overall trend of the other centralities. A similar effect is observed in the centrality dependence of $S\sigma$, which tends to increase smoothly from peripheral to mid-central collisions, where it becomes almost constant. The NBD baseline is in good agreeement with the centrality dependence of $S\sigma$ for both net-charge and net-proton results from HIJING.

The NBD baseline is also larger than the measured $\kappa\sigma^{2}$ from HIJING. One possible explanation for this difference is that particle production in HIJING does not always follow a NBD, but contains additional contributions. This demonstrates that though the mean and variance of the separate positive and negative multiplicity distributions are used to calculate the NBD/BD baseline, it does not automatically describe the higher moments of the net-charge (or net-proton) distributions. Though the NBD baseline is larger in magnitude, the qualitative trend of $\kappa\sigma^{2}$ for the net-charge distributions is reproduced. The NBD baseline should represent the maximum value for the higher moments in the absence of any other correlations.

Figure \ref{NetPKVarCent} shows that $\kappa\sigma^{2}$ ($C_{4}/C_{2}$) for the net-proton distribution from HIJING is also approximately centrality independent and have values between $\approx$1.0-1.5. As before, the Poisson baseline for $\kappa\sigma^{2}$ is always exactly equal to one. Unlike for the net-charge distribution (Figure \ref{NetQKVarCent}), the NBD baseline provides both an excellent quantitative and qualitative description of the centrality dependence of $\kappa\sigma^{2}$ for net-protons. This indicates that the higher moments of the net-proton distribution can be explained by assuming the underlying multiplicity distributions are NBD/BD. The moments products for net-proton distributions are more Poisson-like than for the net-charge distributions simply due to the fact that there are fewer particles counted.

In all cases, the statistical error bars are likely to under estimate the true magnitude of the error. The statistical errors on the variables $S\sigma$ and $\kappa\sigma^{2}$ have a non-trivial dependence on the shape of the underlying distribution. The statistical error is dependent on the variance of these underlying distributions \cite{Delta_Theorem_Error}. Therefore, for the same number of events when the number of particles in the analysis is decreased (e.g. net-protons instead of net-charge) the size of the statistical error also decreases because the variance of the net-proton distribution is narrower than that of the net-charge distribution. As such, this method cannot be a wholly unbiased estimation of the statistical error, particularly for cases where the variance of the underlying distribution is small solely due to limited numbers of accepted particles. The statistical error bars shown in all figures is found using the Delta Theorem \cite{Delta_Theorem_Error}, but can also be calculated using standard methods (sub-event and analytical). They are all quantitatively similar, but do not include effects due to the characteristics of the underlying distribution.
 	
\begin{figure}
\centering
\includegraphics[width=0.60\textwidth]{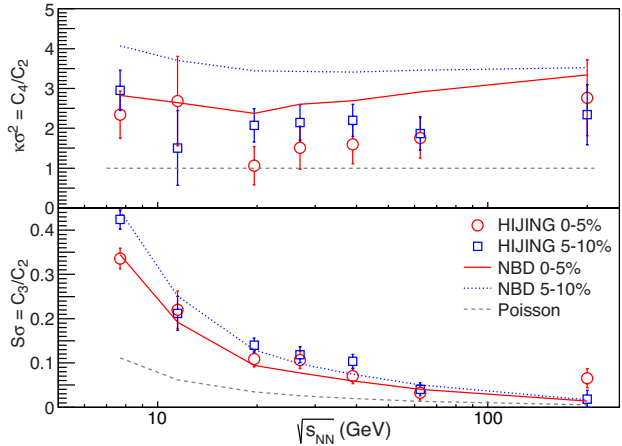}
\caption{Energy dependence of $\kappa\sigma^{2}$ ($C_{4}/C_{2}$) and $S\sigma$ ($C_{3}/C_{2}$) for net-charge distributions from HIJING simulated central (0-5 and 5-10\%) Au+Au collisions. Also shown are predictions assuming the underlying distribution is Poisson (dashed line) or a negative binomial distribution (NBD) (solid line, 0-5\% and dotted line, 5-10\%).}
\label{NetQKVarE}
\end{figure}

\begin{figure}
\centering
\includegraphics[width=0.60\textwidth]{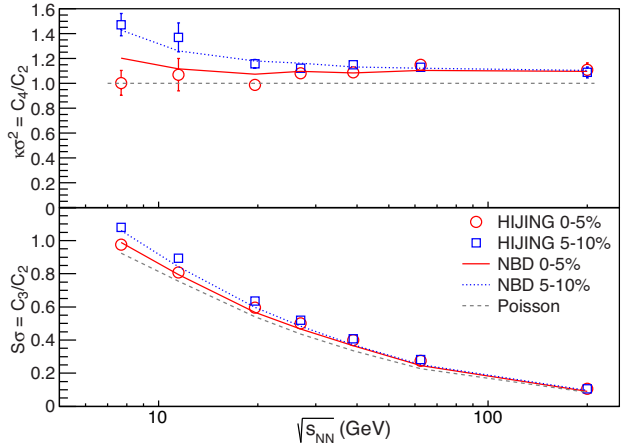}
\caption{Same as Figure \ref{NetQKVarE}, but for net-proton distributions from HIJING simulated central (0-5 and 5-10\%) Au+Au collisions.}
\label{NetPKVarE}
\end{figure}

Figures \ref{NetQKVarE} and \ref{NetPKVarE} shows the incident energy dependence of $S\sigma$ ($C_{3}/C_{2}$) and $\kappa\sigma^{2}$ ($C_{4}/C_{2}$) for HIJING simulated central 0-5 and 5-10\% Au+Au collisions at $\sqrt{\rm s_{NN}}$ = 7.7, 11.5, 19.6, 27, 39, 62.4 and 200 GeV for net-charge and net-proton distributions, respectively. $\kappa\sigma^{2}$ for both the net-charge and net-proton distributions exhibit no overall energy dependence. $S\sigma$ demonstrates a smooth decrease from $\sqrt{\rm s_{NN}}$ = 7.7-200 GeV. For net-charge (net-proton) measurements, $S\sigma$ is intrinsically related to the negative/positive (anti-proton/proton) ratio. Also plotted in Figures \ref{NetQKVarE} and \ref{NetPKVarE} are the Poisson (dashed line) and NBD (solid and dotted lines) baselines for both $S\sigma$ and $\kappa\sigma^{2}$. It is clear from the comparison that the NBD provides a better description of the energy dependence of $S\sigma$ and $\kappa\sigma^{2}$ than the Poisson baseline for both net-charge and net-proton distributions. Some of the values of $\kappa\sigma^{2}$ for 0-5\% most central HIJING events do fall below the NBD baseline. 
Comparing the 0-5\% and 5-10\% bin (which should nominally contain the same physics) in Figures \ref{NetQKVarCent} and \ref{NetPKVarCent} it is shown that the 5-10\% bin does not exhibit these strong excursions from the overall centrality trend.

%

\section{Summary and Conclusion}

The centrality and energy dependence of the moments (mean, variance, skewness, and kurtosis) and their products ($\kappa\sigma^{2}$ and $S\sigma$) of net-charge and net-proton multiplicity distributions in Au+Au collisions simulated by HIJING at $\sqrt{\rm s_{NN}}$ = 7.7, 11.5, 19.6, 27, 39, 62.4, and 200 GeV have been studied. These results have been compared to two cases where the underlying multiplicity distributions are assumed to be either a Poisson (uncorrelated particle production) or a negative binomial/binomial distribution (NBD/BD). The negative binomial distribution has been shown to accurately reproduce features of multiplicity distributions in high-energy elementary ($pp$) and nucleus-nucleus collisions. It has been demonstrated that the Poisson is a poor approximation of the multiplicity distribution for both positive (and negative) particles or protons (and anti-protons). For the first time it is shown that the NBD/BD is a better approximation of the underlying multiplicity distributions and provides a comprehensive description for the main centrality and energy dependence trends of the moments products $\kappa\sigma^{2}$ and $S\sigma$. This insight indicates that relatively small deviations from the statistical Poisson assumption, which are well described by the NBD/BD assumption, cannot provide an unambiguous indication that $\kappa\sigma^{2}$ and $S\sigma$ are sensitive to the quark/gluon-hadron phase transition or any potential QCD critical point.



\bibliography{all}
\bibliographystyle{apsrev4-1}

\end{document}